\documentclass[a4paper, reprint, showkeys, nofootinbib, oneside,superscriptaddress]{revtex4-1}
\usepackage{graphicx}
\usepackage{lipsum} 
\usepackage{color}
\usepackage[english]{babel}
\usepackage[utf8]{inputenc}
\usepackage[colorinlistoftodos, color=green!40, prependcaption]{todonotes}
\usepackage{amsthm}
\usepackage{mathtools}
\usepackage{physics}
\usepackage{xcolor}
\usepackage{graphicx}
\usepackage[left=23mm,right=13mm,top=35mm,columnsep=15pt]{geometry}
\usepackage{adjustbox}
\usepackage{placeins}
\usepackage[T1]{fontenc}
\usepackage{lipsum}
\usepackage{csquotes}

\usepackage{hyperref}
\usepackage{amsmath}
\usepackage{amsfonts}
\usepackage{amssymb}
\usepackage{enumitem}
\usepackage{float}
\usepackage{mathtools}
\usepackage{comment}
\usepackage{capt-of}
\usepackage{placeins} 

\usepackage{adjustbox}



\makeatletter
\renewcommand{\eqref}[1]{(\ref{#1})}
\makeatother

\newcommand{\ctext}[1]{\raise0.2ex\hbox{\textcircled{\scriptsize{#1}}}}

\begin{document}

\title{Jiggled interferometer: Ground-based gravitational wave detector using rapidly-repeated free-falling test masses}

\author{Shoki Iwaguchi}
\email{iwaguchi_s@u.phys.nagoya-u.ac.jp}
\affiliation{Department of Physics, Nagoya University, Furo-cho, Chikusa-ku, Nagoya, Aichi 464-8602, Japan}

\author{Bin Wu}
\affiliation{Department of Physics, Syracuse University, Syracuse, New York 13244, USA}

\author{Kurumi Umemura}
\affiliation{Department of Physics, Nagoya University, Furo-cho, Chikusa-ku, Nagoya, Aichi 464-8602, Japan}

\author{Tomohiro Ishikawa}
\affiliation{Department of Physics, Nagoya University, Furo-cho, Chikusa-ku, Nagoya, Aichi 464-8602, Japan}

\author{Kenji Tsuji}
\affiliation{Department of Physics, Nagoya University, Furo-cho, Chikusa-ku, Nagoya, Aichi 464-8602, Japan}

\author{Ryota Nishimura}
\affiliation{Equipment Development Support section, Technical Center of Nagoya University, Nagoya, Aichi 464-8602, Japan}

\author{Yuta Michimura}
\affiliation{Research Center for the Early Universe (RESCEU), Graduate School of Science, The University of Tokyo, Bunkyo-ku, Tokyo 113-0033, Japan}
\affiliation{Kavli Institute for the Physics and Mathematics of the Universe (Kavli IPMU), WPI, UTIAS, The University of Tokyo, Kashiwa, Chiba 277-8568, Japan}

\author{Yutaro Enomoto}
\affiliation{Institute of Space and Astronautical Science, Japan Aerospace Exploration Agency, Sagamihara, Kanagawa 252-5210, Japan}

\author{Soichiro Morisaki}
\affiliation{Institute for Cosmic Ray Research, The University of Tokyo, 5-1-5 Kashiwanoha, Kashiwa, Chiba 277-8582, Japan}

\author{Yoichi Aso}
\affiliation{National Astronomical Observatory of Japan, Mitaka, Tokyo 181-8588, Japan}

\author{Tomotada Akutsu}
\affiliation{National Astronomical Observatory of Japan, Mitaka, Tokyo 181-8588, Japan}

\author{Keiko Kokeyama}
\affiliation{Department of Physics, Nagoya University, Furo-cho, Chikusa-ku, Nagoya, Aichi 464-8602, Japan}
\affiliation{The Kobayashi-Maskawa Institute for the Origin of Particles and the Universe, Nagoya University, Nagoya, Aichi 464-8602, Japan}
\affiliation{School of Physics and Astronomy, Cardiff University, Cardiff CF24 3AA, United Kingdom}

\author{Seiji Kawamura}
\affiliation{Department of Physics, Nagoya University, Furo-cho, Chikusa-ku, Nagoya, Aichi 464-8602, Japan}

\begin{abstract}
\noindent
We propose the Jiggled Interferometer (JIGI), a novel ground-based gravitational wave detector employing low-frequency noise mitigation similar to that of space-based detectors. Using rapidly-repeated free-fall test masses, JIGI eliminates seismic and suspension thermal noise during free fall. Compared to the Juggled Interferometer, it offers improved angular stability and avoids tracking lasers. We analyze detrending---a required step to remove actuation-induced noise—and show sensitivity gains of about four orders of magnitude in the 0.1–0.3 Hz band, relative to seismic and suspension noise extrapolated from the Cosmic Explorer.
\end{abstract}

\maketitle

Current ground-based gravitational wave (GW) detectors, such as LIGO \cite{LIGO_ref}, Virgo \cite{Virgo_ref}, and KAGRA \cite{KAGRA_ref}, are sensitive to signals at frequencies as low as $\sim$ 10 Hz, primarily from compact binary coalescences. Expanding GW astronomy requires broader frequency coverage and improved sensitivity. Next-generation detectors—including the Einstein Telescope (ET) \cite{ET} and the Cosmic Explorer (CE) \cite{CE}—aim to reach $\sim$ 3 Hz \cite{ET_LF_limit} and $\sim$ 5 Hz \cite{CE_Design}, respectively. However, the regime below a few Hz also contains a rich variety of sources, including intermediate-mass black hole mergers \cite{IMBH_GW}, gravitational waves from primordial black holes \cite{PBH_GW}, and primordial GWs \cite{PGW}. Achieving high sensitivity in this band remains challenging due to the dominant seismic and suspension thermal noise \cite{ET_Cryo, Virgo_ET_Control, ET_SAS, LIGOLF}.

One approach to overcoming low-frequency limitations is the use of space-based detectors, such as the Laser Interferometer Space Antenna (LISA) \cite{LISA} and the Deci-hertz Interferometer Gravitational Wave Observatory (DECIGO) \cite{DECIGO_1, DECIGO_2}. In space, test masses can remain in continuous free fall, avoiding seismic and suspension thermal noise entirely, which significantly enhances sensitivity below 10 Hz. However, despite these advantages, space-based observatories face substantial challenges, including high development costs, complex logistics, and long deployment timelines. In contrast, ground-based detectors offer practical advantages: they are accessible for maintenance, allow for regular calibration and upgrades, and support iterative design improvements throughout their operational lifetime—capabilities that are difficult or impossible in space-based missions. 

As a hybrid approach that aims to combine the benefits of both ground-based and space-based detectors, the Juggled Interferometer (JIFO) has been proposed \cite{JIFO}. This concept seeks to replicate the free-fall environment of space within a terrestrial setup, thereby eliminating seismic and suspension thermal noise while maintaining accessibility and upgradeability. The core idea of the JIFO is to implement a test mass in periodic free fall, actuated by a system such as a linear stage. Each free-fall event typically lasts about one second, corresponding to a height of approximately one meter. This design allows the JIFO to suppress ground-induced noise much like a space-based detector. As a result, its sensitivity improves significantly at frequencies below 10 Hz. However, two major experimental challenges remain: angular instability of the test mass during free fall reduces interference contrast, and the laser system must precisely track the test mass's motion.

\begin{figure}[H]
    \centering
    \includegraphics[clip,width=0.3\textwidth]{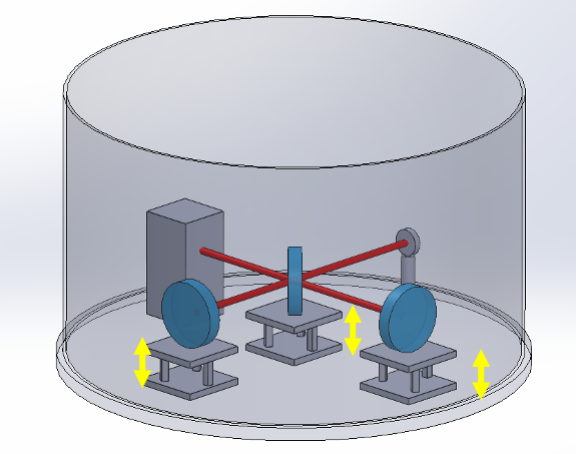}
    \caption{Conceptual design of the JIGI.}
    \label{fig:1}
\end{figure}

To address the limitations of the JIFO, we propose a novel interferometric design that employs rapidly-repeated free-falling test masses on the ground, termed the Jiggled Interferometer (JIGI). The conceptual design of the JIGI is illustrated in Fig. \ref{fig:1}.

Like the JIFO, the JIGI retains the noise-mitigation benefits of space-based detectors while preserving the flexibility and accessibility inherent to ground-based setups. Distinctively, the JIGI significantly shortens the duration of each free-fall—typically to 0.01 seconds, corresponding to a fall distance of approximately 0.1 mm—allowing the test masses to be actuated using piezoelectric transducers.

The shortened free-fall duration provides two key advantages: improved angular stability of the test masses and the elimination of laser tracking requirements. To maintain high interference fringe contrast, the mirrors must remain angularly stable during free fall. For a given initial angular velocity, a shorter fall time results in smaller angular deviations, making the JIGI inherently more stable than the JIFO. As for laser tracking, while fiber-based tracking systems are theoretically viable, they tend to introduce additional noise. In proof-of-principle experiments with the JIFO, both angular instability and fiber-optic tracking presented substantial technical difficulties. In contrast, the minimal displacement in the JIGI removes the need for active tracking, and the angular stability is intrinsically enhanced. These improvements make the JIGI not simply a scaled-down version of the JIFO, but a qualitatively different architecture with significantly reduced implementation challenges. Table \ref{tab:JIFO_JIGI} summarizes the key differences between JIFO and JIGI.

\begin{table*}[t]
\centering
\caption{Comparison of JIFO and JIGI.}
\label{tab:JIFO_JIGI}

\setlength{\tabcolsep}{8pt}  
\renewcommand{\arraystretch}{1.1} 

\begin{tabular}{lccc}
\hline\hline
Interferometer type & Free-fall time & Tracking laser & Angular stability of test mass \\
\hline
JIFO & $\sim 1\,\mathrm{s}$    & Required     & Problematic \\
JIGI & $\sim 0.01\,\mathrm{s}$ & Not required & Robust \\
\hline\hline
\end{tabular}
\end{table*}

Notably, in both the JIGI and the JIFO, data can only be acquired during the brief intervals of free fall. To achieve continuous observation, at least two interferometers must operate in an alternating sequence, ensuring that one is always in free fall while the other resets. In the remainder of this letter, we assume such an alternating configuration to enable uninterrupted data acquisition.

\begin{figure}[H]
    \centering
    \includegraphics[clip,width=0.35\textwidth]{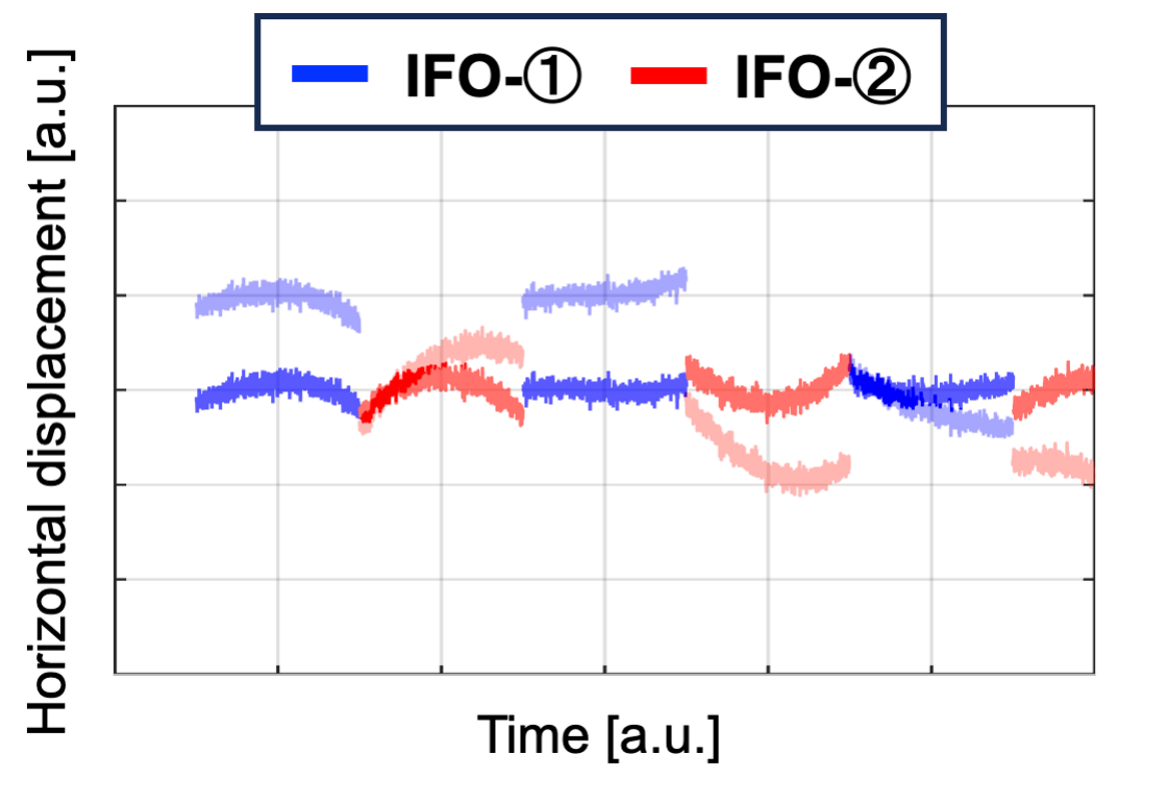}
    \caption{Comparison of the horizontal displacement signal including GW signal and white noise, before and after detrending. The light and dark-colored curves show the horizontal displacement signal before and after detrending, respectively. The red and blue curves represent the signals from two interferometers operating alternately, respectively. All quantities are shown in arbitrary units for methodological illustration and do not represent actual detector displacement limits.}
    \label{fig:2}
\end{figure}

In interferometers employing repeatedly free-falling test masses, the horizontal displacement signal includes random zeroth- and first-order components—specifically, the initial position and velocity of the test mass. The displacement signal x(t) can be modeled as: $x(t) = h(t) + n(t) + (v_\mathrm{initial}t + x_\mathrm{initial})$, where h(t) is the GW signal, n(t) represents noise, $v_\mathrm{initial}$ is the initial velocity, and $x_\mathrm{initial}$ is the initial displacement at the moment of release. These deterministic terms are introduced by the actuation process and act as noise in the measurement. To remove them, a detrending method is applied during post-processing, in which a linear fit is subtracted from the displacement signal. The detrended signal is given by $s_\mathrm{detrend} = x(t) - (at+b)$, where $a$ and $b$ are the coefficients of the linear and constant components, respectively. As shown in Fig. \ref{fig:2}, this procedure effectively eliminates the actuation-induced trends.

Figure \ref{fig:2} shows the displacement signal including GW signal and white noise, before and after detrending. The horizontal displacement signal before detrending contains random zeroth-order and first-order terms for each free-falling cycle, indicated by the light-colored curves in Fig. \ref{fig:2}. In this letter, the random components, $v_\mathrm{initial}t + x_\mathrm{initial}$, and the linear fit term, $at+b$, are referred to as the random trend and the fit trend, respectively.

\begin{figure}[H]
    \centering
    \includegraphics[clip,width=0.45\textwidth]{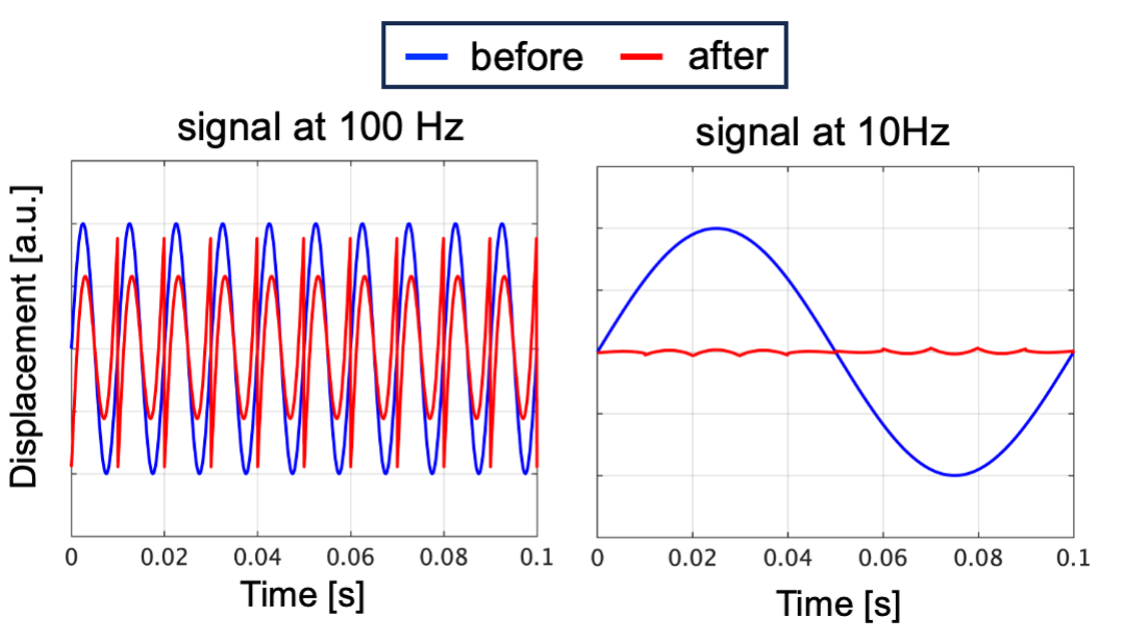}
    \caption{Comparison of  100 Hz and 10 Hz GW signals (simplified to sine waves) before and after detrending, with the jiggling frequency of 100 Hz. The vertical axis is shown in arbitrary units for methodological illustration and does not represent actual detector displacement limits.}
    \label{fig:3}
\end{figure}

While the detrending method effectively removes the random trend, it also alters the gravitational wave signal. Previous studies \cite{JIFO_Bin} have shown that detrending suppresses signal components at frequencies lower than the inverse of the free-fall duration. This suppression arises because the linear fit approximates not only the unwanted trends but also part of the GW signal, particularly its low-frequency content. As illustrated in Fig. \ref{fig:3}, gravitational wave signals at 10 Hz and 100 Hz are both affected by the detrending process, with the lower-frequency signal experiencing significantly greater attenuation.

Although GW signal is suppressed by the detrending method, noise is also reduced in the same proportion, so the signal-to-noise ratio (SNR) is preserved before and after detrending. However, considering the full frequency spectrum, the SNR is not preserved. While previous studies discussed the SNR at a single frequency, the discussion in this letter considers all frequency components. In this situation, the detrended signal is subject to the aliasing. Aliasing occurs because the process of obtaining the fit trend can be regarded as a form of sampling, with the effective sampling frequency determined by the duration of the free fall.

To evaluate the frequency-dependent effects of detrending, we analyze the fit trend with a least-squares approach. For simplicity, consider a sinusoidal signal

\begin{equation}
  s(t) = A e^{i \omega_0 t} ,
\notag
\end{equation}

\noindent
where $A$ is the amplitude and $\omega_0$ the angular frequency. For the $k$-th free-fall interval $(k-1)T_j \leq t < kT_j$, with duration $T_j$, the squared fitting error is

\begin{equation}
  E_k(a_k,b_k)
  = \int_{(k-1)T_j}^{kT_j} \bigl[s(t) - (a_k t + b_k)\bigr]^2 \, dt ,
\notag
\end{equation}

\noindent
where $a_k$ and $b_k$ are the slope and intercept of the linear fit. Minimizing $E_k$ yields the least-squares conditions:

\begin{align}
\frac{\partial E_k}{\partial a_k}
&= -2 \int_{(k-1)T_j}^{kT_j} t \bigl[\, s(t) - (a_k t + b_k) \,\bigr] \, dt = 0,
\notag \\
\frac{\partial E_k}{\partial b_k}
&= -2 \int_{(k-1)T_j}^{kT_j} \bigl[\, s(t) - (a_k t + b_k) \,\bigr] \, dt = 0,
\notag
\end{align}

Solving yields the slope $a_k$ and intercept $b_k$ of the linear fit:

\begin{equation}
\begin{aligned}
a_k &= \frac{I_{k,t}\, I_{k,s} - I_{k,0}\, I_{k,ts}}{I_{k,t}^{2} - I_{k,0}\, I_{k,t^{2}}}, 
b_k &= \frac{I_{k,t^{2}}\, I_{k,s} - I_{k,t}\, I_{k,ts}}{I_{k,t^{2}}\, I_{k,0} - I_{k,t}^{2}} ,
\end{aligned}
\notag
\end{equation}

\noindent
where

\begin{equation}
\begin{aligned}
I_{k,0}   &= \int_{(k-1)T_j}^{kT_j} dt,         &
I_{k,t}   &= \int_{(k-1)T_j}^{kT_j} t\, dt, \\
I_{k,t^2} &= \int_{(k-1)T_j}^{kT_j} t^2\, dt,   &
I_{k,s}   &= \int_{(k-1)T_j}^{kT_j} s(t)\, dt, \\
I_{k,ts}  &= \int_{(k-1)T_j}^{kT_j} t\, s(t)\, dt.
\end{aligned}
\notag
\end{equation}

\noindent
These coefficients define the linear fit term $a_k t + b_k$ for each segment. The detrended signal is

\begin{equation}
  s_{\mathrm{detrend}}(t) = s(t) - s_{\mathrm{trend}}(t) ,
    \notag
\end{equation}

For a monochromatic input $s(t)=Ae^{i\omega_0 t}$, the Fourier spectrum of the trend shows suppression around $\omega_0$ and aliasing from segmentation. Because detrending is applied to each free-fall segment, it acts like sampling with effective rate $1/T_j$. The detrended signal therefore exhibits aliasing: frequency components are down-converted into lower bands. In particular,




\[
\omega = n \omega_j \pm \omega_0, \quad n \in \mathbb{Z}, \quad \omega_j = \frac{2\pi}{T_j},
\]

\noindent
appear in the detrended spectrum even if absent in the original signal. Thus SNR is not uniformly preserved, and degrades below $\omega_j$ where aliasing dominates.

To quantify aliasing, consider $s(t) = A e^{i \omega_0 t}$. The Fourier amplitude of the fit trend is

\begin{equation}
  S_{\mathrm{trend}}(\omega,\omega_0)
  = \int_{-\infty}^{\infty} s_{\mathrm{trend}}(t)\, e^{i\omega t}\, dt .
\notag
\end{equation}

\noindent
For $N$ equal segments of duration $T_j$, one finds

\begin{equation}
  \begin{aligned}
    S_{\mathrm{trend}}(\omega,\omega_0)
    &= \frac{2A}{\omega^{2}\,\omega_{0}^{2}\,T_{j}^{4}\,N}
       \cdot \frac{e^{i(\omega_{0}-\omega) N T_{j}} - 1}{e^{i(\omega_{0}-\omega) T_{j}} - 1}
       \cdot F(\omega_{0}, \omega_{j}) .
  \end{aligned}
\notag
\end{equation}

\noindent
where $F(\omega_0, \omega_j)$ encodes the dependence on the jiggling frequency. Only frequencies $\omega = n \omega_j \pm \omega_0$ survive, confirming that detrending redistributes power to harmonics of $\omega_j$.

The Fourier component of the detrended signal at $\omega = \omega_0$ has two parts: (1) a direct term from the fit to $\omega_0$, and (2) aliasing terms from $n\omega_j \pm \omega_0$.  

For $\omega_0 T_j \ll 1$, the direct term scales as

\[
\sim A f_0^4 T_j^4, \qquad f_0 = \frac{\omega_0}{2\pi},
\]
while aliasing scales as
\[
\sim A f_0^2 T_j^2.
\]

\noindent
Thus without aliasing, amplitude is suppressed by $f_0^4 T_j^4$; with aliasing, suppression weakens to $f_0^2 T_j^2$, so aliased signals dominate at low frequency. Consequently, SNR is not preserved below $f_j = 1/T_j$ because aliasing shifts noise power into that band.

\begin{figure}[H]
    \centering
    \includegraphics[clip,width=0.4\textwidth]{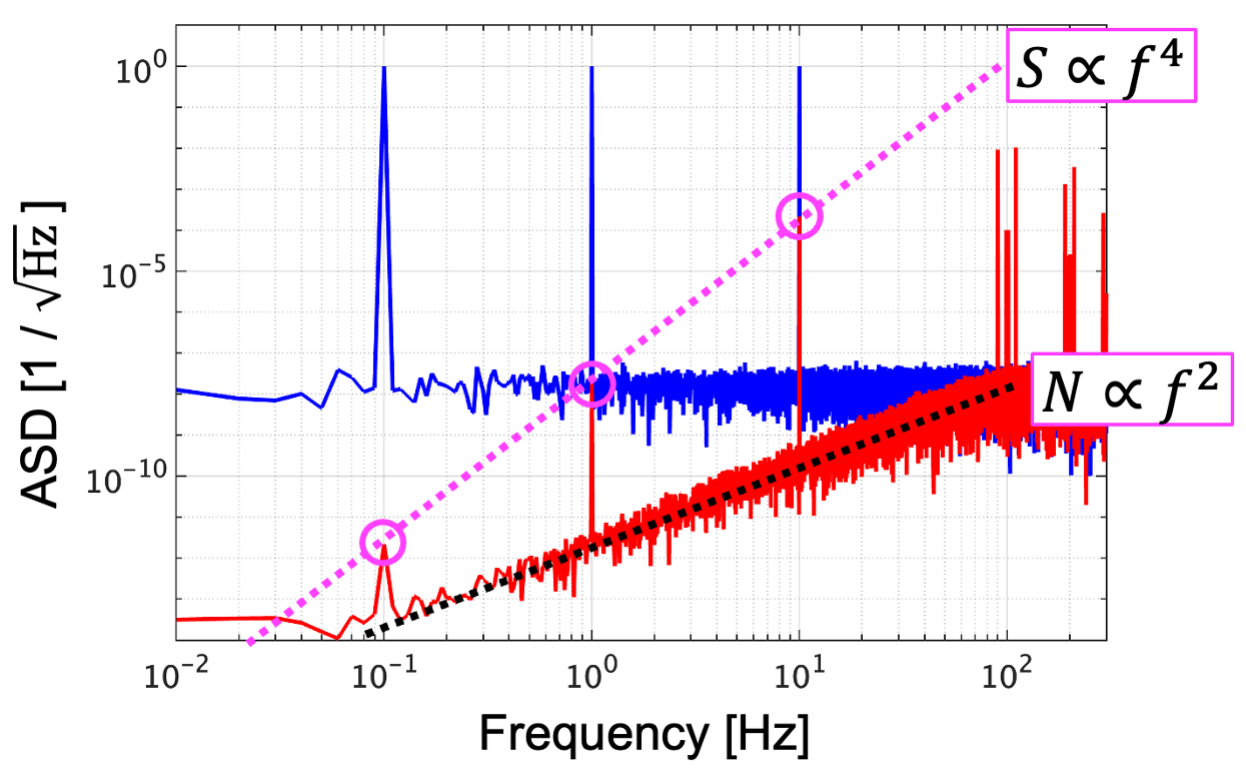}
    \caption{Spectrum of the signal including the GW signals at 0.1 Hz, 1 Hz, and 10 Hz, and white noise, before (blue) and after (red) detrending, with the jiggling frequency of 100 Hz. The magenta and black dashed curves indicate the frequency dependence of the reduction ratio below $f_\mathrm{j}$ for the GW signals and white noise, respectively. This figure verifies the $f^2$ reduction ratio using a monochromatic test signal and does not represent the final detector noise floor.}
    \label{fig:4}
\end{figure}

For a more comprehensive explanation, we discuss the spectrum of the signal including the GW signal and white noise, as shown in Fig. \ref{fig:4}.

To illustrate the spectral effects of detrending, we simulate a composite signal containing GW components at 0.1, 1, and 10 Hz, plus white noise. Figure. \ref{fig:4} shows the amplitude spectral density (ASD) before and after detrending with a jiggling frequency $f_\mathrm{j}$ = 100 Hz. The white noise is generated independently for each segment. In Fig. \ref{fig:4}, the dashed curves indicate the frequency dependence of monochromatic signals and white noise in the reduction ratio before and after detrending below $f_\mathrm{j}$, showing that the SNR decreases proportionally to $f^2$ at frequencies below $f_\mathrm{j}$. This result means that the noise dominant around $f_\mathrm{j}$ is down-converted to lower frequencies, with its contribution scaling as $f^{-2}$.


Several strategies have been considered to mitigate aliasing, but each has limitations. First, reducing noise around $f_\mathrm{j}$ before detrending would lower down-converted contributions, but pre-filtering distorts each segment's trend, making linear subtraction ineffective. Second, reconstructing the original signal by slightly overlapping adjacent data segments might seem effective. In practice, however, this fails because the noise in each segment is statistically independent, making accurate reconstruction impossible. Third, one might attempt to recover the original signal at relevant frequencies by applying the inverse of the matrix associated with the detrending operation. Unfortunately, this matrix is not invertible, rendering the method infeasible. Finally, matched filtering can suppress some aliased noise, but its effectiveness is limited by noise near $f_\mathrm{j}$. More generally, neither pre-processing nor post-processing offers a complete mitigation: pre-filtering that suppresses noise near $f_\mathrm{j}$ distorts the intra-segment trend itself and thereby makes detrending ineffective, while detrending is a non-invertible projection at the segment level, so the associated SNR loss cannot be recovered by post-processing alone.

\begin{figure}[H]
    \centering
    \includegraphics[clip,width=0.45\textwidth]{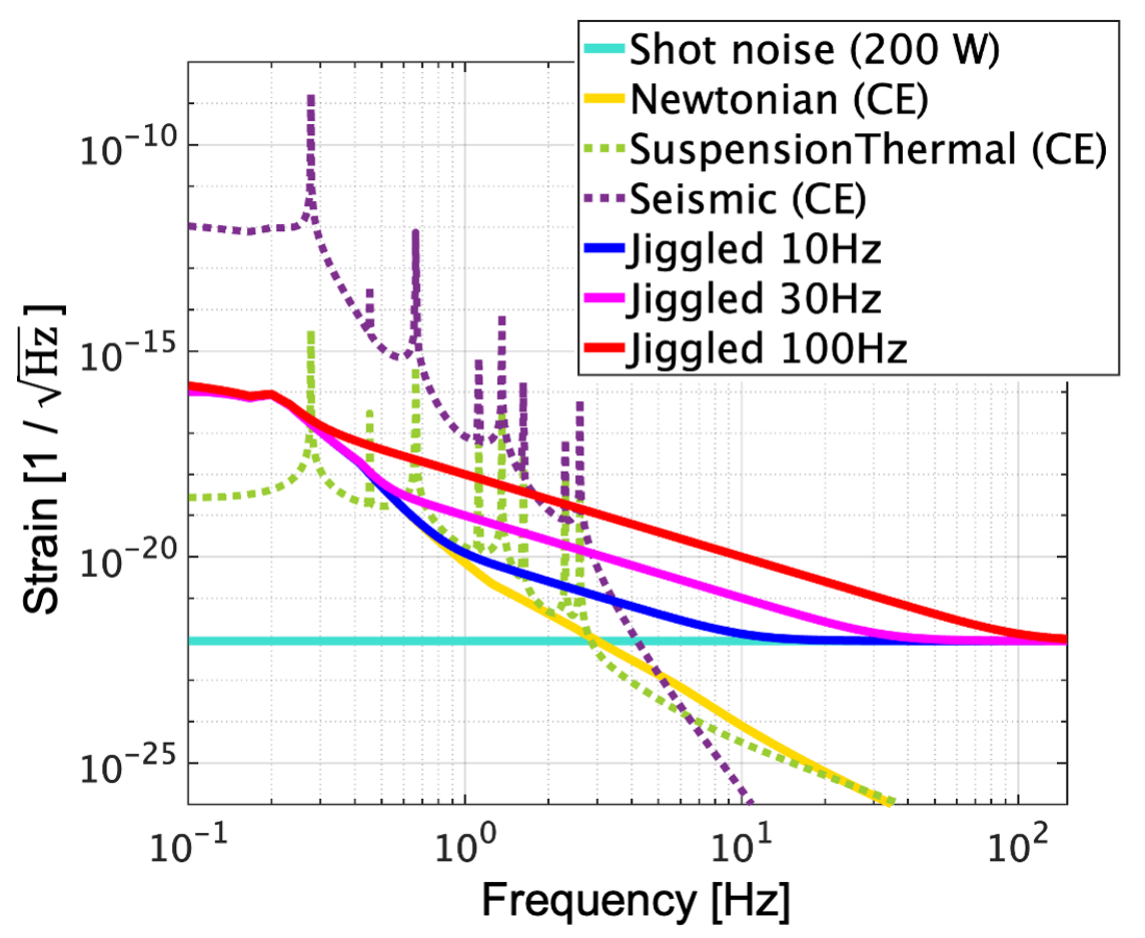}
    \caption{Estimated sensitivity of the JIGI along with noise budgets. The cyan curve represents the shot noise. The Newtonian noise, seismic noise, and suspension thermal noise for the CE are shown in yellow, purple, and green, respectively. The CE noise budgets were extrapolated using pygwinc \cite{pygwinc}. The red, magenta, and blue curves show the total sensitivity of the JIGI, including down-converted noise, for jiggling frequencies of 10 Hz, 30 Hz, and 100 Hz, respectively.}
    \label{fig:5}
\end{figure}

We now translate the above aliasing analysis into an equivalent strain sensitivity to assess the overall detector performance. Figure \ref{fig:5} presents estimated JIGI sensitivity, including relevant noise sources. The configuration assumes a 40 km Michelson interferometer with 200 W of laser power; shot noise is calculated accordingly. Of the dominant low-frequency noise sources in Cosmic Explorer (CE)—seismic, suspension thermal, and Newtonian—only Newtonian noise is relevant to JIGI. Seismic and suspension noise are absent because free fall eliminates coupling to ground and suspensions. CE curves in Fig. \ref{fig:5} are extrapolated using pygwinc \cite{pygwinc}; see \cite{CE_LF_limit} for CE low-frequency limits. Assuming shot noise dominates near $f_\mathrm{j}$, the down-converted noise scales as $f^{-2}$ below $f_\mathrm{j}$. Thus, JIGI’s sensitivity is limited by Newtonian noise, shot noise, and aliasing. Even so, JIGI achieves far better sensitivity below 1 Hz than CE, with ~4 orders of magnitude improvement in the 0.1–0.3 Hz band. With a jiggling frequency of $30\,\mathrm{Hz}$, JIGI reaches a strain sensitivity of $\sim 1 \times 10^{-19}\,\mathrm{Hz}^{-1/2}$ near $1\,\mathrm{Hz}$. At this sensitivity, quasinormal-mode signals from a massive black-hole merger can, for example, be detected with an SNR of $10$ for a source at a luminosity distance of $\sim 0.2\,\mathrm{Gpc}$~\cite{JIFO_Bin}, corresponding to a remnant black-hole mass of $\sim 2.6 \times 10^{4}\,M_{\odot}$. Current astrophysical estimates for massive black-hole ($10^{3}$--$10^{5}\,M_{\odot}$) merger rates span a broad range, typically $0.01$--$1\,\mathrm{Gpc}^{-3}\,\mathrm{yr}^{-1}$~\cite{Fragione_GC,Rasskazov_GC,Fragione_NSC}, implying a low but non-zero expected rate within $0.2\,\mathrm{Gpc}$. If the vertical beam-shift limitation can be mitigated, the noise level near $1\,\mathrm{Hz}$ could improve by roughly one order of magnitude (for a jiggling frequency of $10\,\mathrm{Hz}$), increasing the accessible volume by $\sim 10^{3}$ and substantially enhancing the detection prospects.

In the JIGI, while the laser remains fixed, the test mass undergoes vertical motion in each free-fall cycle, shifting the beam spot on the mirror. To avoid degraded optical performance, vertical motion must be limited to ~1 mm, imposing a lower bound of ~30 Hz on $f_\mathrm{j}$. Below this, beam spot deviation risks coupling into higher-order modes or causing clipping, introducing excess noise and reducing contrast.

Sensitivity may be further improved with Fabry–Perot cavities and quantum squeezing. A Fabry–Perot cavity enhances the displacement signal, effectively reducing shot noise, while squeezed vacuum states suppress quantum noise, both near $f_\mathrm{j}$. Both are technically demanding for JIGI: cavities require stable alignment in a rapidly actuated system, and squeezing must withstand changing optical paths. These techniques could enhance sensitivity, but their integration into JIGI remains challenging.


To achieve the estimated sensitivity, we consider two practical constraints. First, vertical motion couples into the longitudinal direction due to Earth’s curvature. For a 40 km interferometer, a 1 mm vertical displacement induces ~1 $\mathrm{\mu}$m longitudinal motion. Because free-fall trajectories are not perfectly linear, detrending cannot fully remove this coupling. Although this coupling appears primarily at the jiggling frequency and, in principle, does not affect lower frequencies, variations in the jiggling frequency due to timing errors can cause aliasing into lower frequencies. Therefore, stable timing of the analysis segment boundaries is required; this concerns data time-stamping rather than mechanical reproducibility of the free-fall motion. Second, actuation forces may excite structural resonances, which down-convert into the observation band. To mitigate this excitation, the actuation stress should be optimized to suppress structural resonances. Furthermore, the impact of this aliasing is attenuated by the square of the ratio between the aliased and original frequencies, allowing resonances to be considered negligible in practice. For example, a 1010 Hz resonance aliased to 10 Hz at $f_\mathrm{j}$ = 100 Hz is attenuated by $10^{-4}$.

In summary, we propose JIGI—a novel GW detector for high sensitivity below a few hertz. By employing rapidly-repeated free falls, JIGI eliminates seismic and suspension thermal noise. While conceptually related to the Juggled Interferometer (JIFO), JIGI uses much shorter free falls, improving angular stability and avoiding active laser tracking. We also analyze detrending, which, while necessary to suppress actuation-induced noise, introduces aliasing that redistributes power across the spectrum. Despite this, JIGI maintains superior sensitivity below a few Hz compared to extrapolated CE noise. These results establish JIGI as a promising architecture for next-generation low-frequency GW detection.

To further assess the feasibility of this approach, we are currently conducting a proof-of-principle experiment to test the core operating mechanism of the JIGI. Preliminary measurements demonstrate the feasibility of achieving repeatable sub-millimeter free-fall with high angular stability, and detailed experimental results will be reported in a forthcoming publication. While still in progress, this effort aims to provide initial validation of the rapidly-repeated free-fall technique and guide future development. Successful validation will position the JIGI as a promising and innovative candidate for next-generation low-frequency GW detection in terrestrial detectors, potentially replacing con- ventional vibration isolation systems.

\vspace{10pt}
\textit{Acknowledgments}\text{---} We thank Kushal Jain for assistance with English editing. We would like to thank Tetsuya Shiromizu, Shintarou Yanagida, and Leo Tsukada for useful discussions. This work was supported by Research Grants from Fujikura Foundation, Sumitomo Foundation, and the Japan Society for the Promotion of Science (JSPS) KAKENHI (Grant No. 21K18626)

\end{document}